\documentclass[conference,a4paper]{IEEEtran}

  	\usepackage[pdftex]{graphicx}
        \usepackage[tight,footnotesize]{subfigure}
  	\graphicspath{{../pdf/}{../jpeg/}}
	\DeclareGraphicsExtensions{.pdf,.jpeg,.png}
	\usepackage[cmex10]{amsmath}
	\usepackage{mathabx}
	\usepackage{algorithmic}
	\usepackage{array}
        \usepackage{tabularx}
	\usepackage{mdwmath}
	\usepackage{mdwtab}
	\usepackage{eqparbox}
	\usepackage{url}
        \usepackage{romannum}
        \usepackage{xcolor}
        \usepackage{balance}
        \usepackage{cite}
\begin{document}


\title{Efficient Ray-Tracing Channel Emulation in Industrial Environments: An Analysis of Propagation Model Impact}

\author{Gurjot Singh Bhatia, Yoann Corre}

\author{\IEEEauthorblockN{
Gurjot Singh Bhatia\IEEEauthorrefmark{1}\IEEEauthorrefmark{2},   
Yoann Corre\IEEEauthorrefmark{1},   
M. Di Renzo\IEEEauthorrefmark{2}}                                     
\IEEEauthorblockA{\IEEEauthorrefmark{1} SIRADEL, Saint-Gregoire, France \\
\IEEEauthorrefmark{2} Universit\'e Paris-Saclay, CNRS, CentraleSup\'elec, Laboratoire des Signaux et Syst\`emes, Gif-sur-Yvette, France\\
gsbhatia@siradel.com}
}

\maketitle

\begin{abstract}
Industrial environments are considered to be severe from the point of view of electromagnetic (EM) wave propagation. When dealing with a wide range of industrial environments and deployment setups, ray-tracing channel emulation can capture many distinctive characteristics of a propagation scenario. Ray-tracing tools often require a detailed and accurate description of the propagation scenario. Consequently, industrial environments composed of complex objects can limit the effectiveness of a ray-tracing tool and lead to computationally intensive simulations. This study analyzes the impact of using different propagation models by evaluating the number of allowed ray path interactions and digital scenario representation for an industrial environment. This study is realized using the Volcano ray-tracing tool at frequencies relevant to 5G industrial networks: 2 GHz (mid-band) and 28 GHz (high-band). This analysis can help in enhancing a ray-tracing tool that relies on a digital representation of the propagation environment to produce deterministic channel models for Indoor Factory (InF) scenarios, which can subsequently be used for industrial network design.


\end{abstract}

\vskip0.5\baselineskip

\IEEEpeerreviewmaketitle


\section{Introduction}

The 5G mobile communication was developed with the goal to not only enhance the broadband capabilities of mobile networks but also supply enhanced wireless access to a wide range of vertical industries, including the industrial, automotive, and agricultural sectors \cite{5g2020key}. Wireless communication plays a pivotal role in the development of modern smart factories, the Industrial Internet-of-Things (IIoT), and Industry 4.0 by providing robust and scalable means of interconnecting various machines, sensors, and mobile entities, such as mobile robots, automated guided vehicles (AGVs), drones, human operators, and other relevant components. With the rapid growth of the IIoT and the increasing demand for real-time communication, it is imperative to understand the communication channel's impact on the system's overall performance. Channel modeling plays a crucial role in ensuring seamless and efficient communication between different devices and systems. Hence, radio channel characterization is an essential topic for radio communication system design.

Propagation in Indoor Factory (InF) scenarios \cite{3gpp2019study} is more site-specific than propagation in usual residential or office environments \cite{sun2016propagation}. Strong non-line-of-sight (NLoS) obstructions and fading fluctuations may cause degraded signal and system reliability. For instance, in InF scenarios, the most common objects in the environment are metallic machines and storage racks. Their huge bodies can become major blockers in the NLoS case \cite{schmieder2020measurement}. The smooth metallic surface creates many reflections, and the object's size hinders the direct transmission of the signal. The InF environment with many such objects complicates the radio propagation in the factory \cite{8377337, schmieder2019directional}.

Ray-tracing is a commonly-used technique in deterministic channel modeling. Its algorithm starts by finding all the possible geometrical rays between a transmitter and a receiver for a given number of allowed interactions in the digital representation of the propagation scenario. Then, the calculation of the rays' (EM field) contributions is based on the geometrical optics (GO), uniform theory of diffraction (UTD), and effective roughness theory (ER), assuming that the far-field conditions are met \cite{indoor}. Ray-tracing can provide path loss data, angle-of-arrival (AoA), angle-of-departure (AoD), time delay, and optical visibility (LoS or NLoS), among other relevant parameters. 

Given the site-specific nature of InF scenarios and the recent interest in millimeter-wave (mmWave) frequency bands for industrial wireless networks, many channel and network performance metrics for mmWave communications have been lately generated through ray-tracing \cite{gougeon2019ray, charbonnier2020calibration}. Ray-tracing can be considered an interesting approach for radio propagation and characterization in InF scenarios.

Ray-tracing is a powerful tool for simulating the propagation of radio waves, but it can require significant computational resources to obtain accurate results, particularly for complex problems. The significance of this factor is further amplified in InF scenarios due to the scale of the industrial setting and the presence of many intricate structures and objects with complex geometries in the environment. It can make ray-tracing simulations computationally intensive, pushing the tool to its limits. 

In this paper, the examination of various propagation models with respect to the maximum number of allowed ray path interactions and digital representation of the industrial scenario in a ray-tracing tool will be carried out. The aim is to determine each propagation model's performance by comparing large-scale channel parameters and global parameters, including computation times (CTs) and root mean square errors (RMSEs).

The results of this analysis will provide insights into the most effective approach for modeling signal propagation in industrial environments. It can help in enhancing the performance of a ray-tracing tool and give valuable perspectives for link- or system-level design and radio-planning for InF scenarios. It can be of particular interest to: 1) Research and Development (R\&D) teams using ray-tracing to virtually test the performance of new technologies or topologies applied to Industry 4.0 use cases; 2) Researchers relying on physical-layer digital twins for developing artificial-intelligence (AI) algorithms; 3) Private network operators or service providers in charge of the radio nodes deployment. To the best of our knowledge, there has yet to be any such attempt for industrial environments.

This paper is structured as follows. Section II describes the propagation scenario and its 3D digital representation. Section III outlines the ray-tracing propagation models that are tested in this analysis. Section IV presents the simulation results and the derived recommendations. Finally, Section V gives the conclusion and future perspective of this work.


\section{Propagation Scenario and Settings}

The propagation scenario chosen for this study is a washing industry used to wash, dry clean, and store large batches of textile products. 

\begin{figure}[ht!] 
\centering
\includegraphics[width=3.3in]{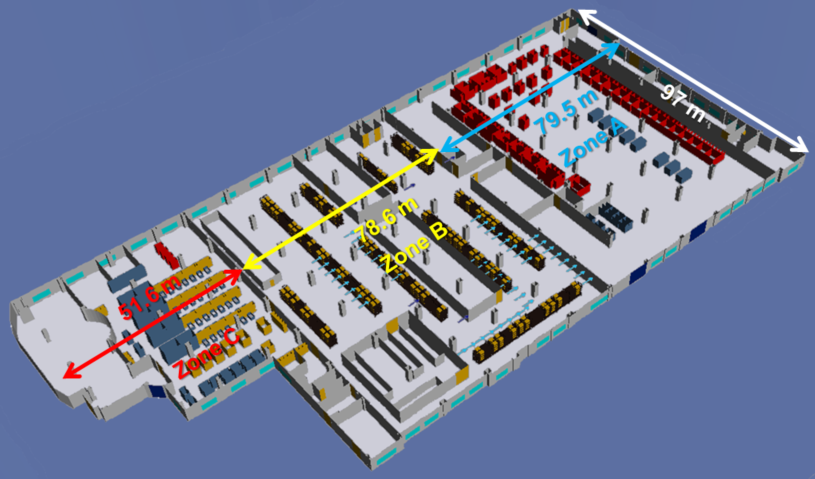}
\caption{3D digital model of the propagation scenario in the channel emulator.}
\label{1.1}
\end{figure}

Fig. \ref{1.1} shows the 3D digital model of the propagation scenario in the channel emulator. The factory floor is 210 m long, 97 m wide, and 5 m high. The factory floor can be divided into three zones. These zones differ from each other in terms of shape, size, object material, and clutter density. Zone A is the biggest. It is 79.5 m long and is characterized by machines and storage containers. Zone B is 78.6 m long, filled with storage racks stacked with wooden boxes. Zone C is 51.6 m long, characterized by metallic lockers, wooden benches, and some metallic housing units.

\begin{figure}[ht!] 
\centering
\includegraphics[width=3.1in]{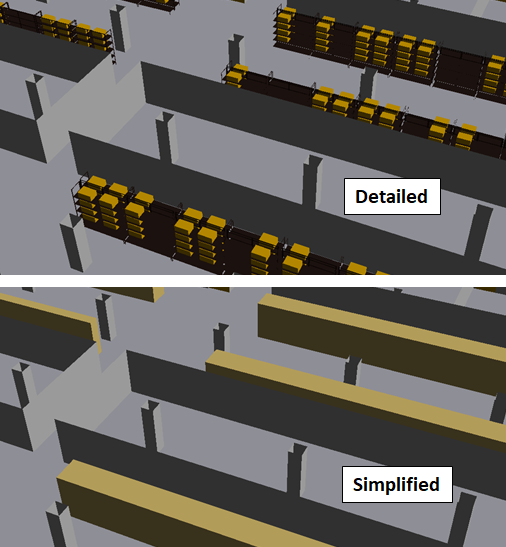}
\caption{Detailed representation (top) vs. Simplified representation (bottom) of the storage racks.}
\label{2.3}
\end{figure}

The site of focus for this study will be Zone B. The impact of metallic storage racks on propagation in Zone B will be analyzed. These racks are 4.4 m in height. Metallic storage racks are one of the most common industrial objects, thus making them a suitable choice for the analysis. As shown in Fig. \ref{2.3} (top), the storage racks are drawn quite in detail. The racks are stacked with wooden boxes of dimensions: 1.2 m long, 0.8 m wide, 0.5 m high, and 20 cm thick. The exact location of those boxes does not correspond to reality but is representative of a storage area filled at around 70\% of its maximum capacity. As shown in Fig. \ref{2.3} (bottom), a simplified scenario is also created where the detailed storage racks are replaced with large cuboids with dimensions similar to envelop the detailed representation. 

\begin{figure}[ht!] 
\centering
\includegraphics[width=3.2in]{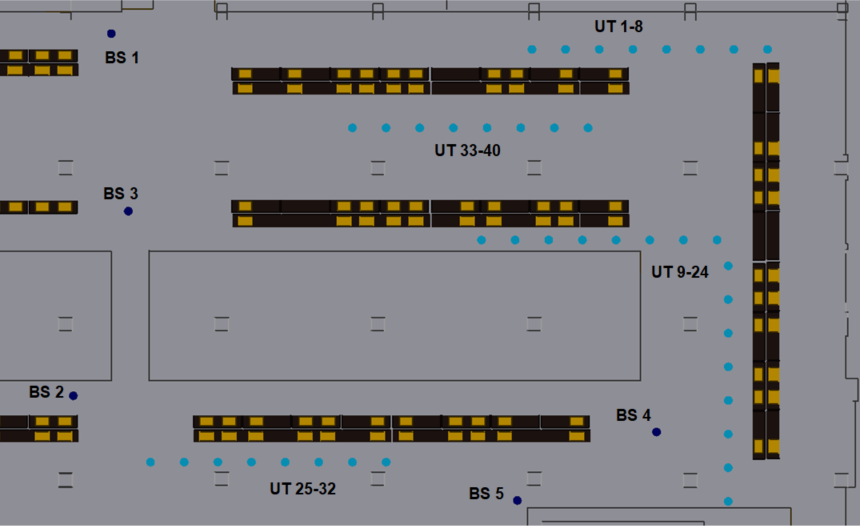}
\caption{Positions of different BSs and UTs.}
\label{3.1}
\end{figure}

There are a total of five base stations (BSs) at different locations that act as transmitters, as shown in Fig. \ref{3.1}. The BS antenna used in simulations is a single-element half-wavelength vertically-polarized omni-directional antenna. There are a total of forty user terminal (UT) positions, which act as receivers. The UT antenna is also a single-element half-wavelength vertically-polarized omni-directional antenna. The BSs are deployed at a height of 4 m above the ground, while the UTs are at a height of 1.5 m above the ground. The analysis is carried out at a carrier frequency of 2 GHz and with a more limited scope at 28 GHz. This design is a virtual test scenario; realistic positions/heights were selected, but this does not correspond to an actual deployment in the factory.


\section{Ray-tracing propagation models under test}

\begin{figure}[ht!] 
\centering
\includegraphics[width=3.3in]{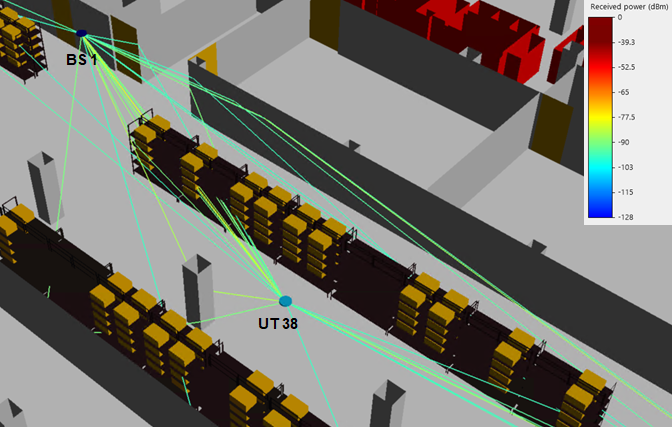}
\caption{Some of the predicted ray paths for point-to-point (P2P) prediction between BS 1 and UT 38, at 2 GHz.}
\label{4.1}
\end{figure}

After creating a digital scenario, point-to-multi-point (P2MP) simulations are realized between different transmitter and receiver positions, using Volcano Flex \cite{Siradel} ray-tracing. Volcano Flex is based on the ray-launching technology. It is a time-efficient ray-based propagation model capable of predicting deterministic path-loss in any small-scale urban or indoor scenario. It can provide channel properties and 3D multi-path trajectories. Fig. \ref{4.1} shows some of the predicted ray paths for one of the links.

In a ray-tracing tool, using quite a detailed representation of the propagation scenario with a high number of allowed interactions would usually be preferred. Consequently, CTs can significantly increase on increasing the number of maximum allowed interactions and the details of the objects in the digital scenario. However, it is not certain that there will be a significant improvement in the accuracy of the predicted channel parameters. Moreover, making detailed objects for a digital scenario can be rather tedious if the propagation environment is extensive, with objects of different shapes and sizes, as in a typical InF scenario. It is crucial to consider real constraints such as CTs and 3D modeling capabilities when producing deterministic ray-based channel models for InF scenarios. Hence, different propagation models with detailed and simplified representations and a varying number of allowed interactions are analyzed in this study:

\begin{enumerate}
    \item Reference model: detailed representation of storage racks. They are associated with the material properties of a typical metal \cite{ITU}. The wooden boxes have material properties of wood \cite{ITU}: $\varepsilon_\mathrm{r}^{'} = 1.99$, $\varepsilon_\mathrm{r}^{''} = 0.090$ at 2 GHz and $\varepsilon_\mathrm{r}^{'} = 1.99$, $\varepsilon_\mathrm{r}^{''} = 0.11$ at 28 GHz, where $\varepsilon_\mathrm{r}^{'}$ and $\varepsilon_\mathrm{r}^{''}$ are real and imaginary part of relative permittivity, respectively. Reflections and diffractions are allowed due to the large horizontal metallic surfaces of the racks and the wooden boxes. A maximum of three successive reflections and one diffraction (3R1D) are allowed for each ray. There is no limitation on the number of transmissions. 
    \item 2R1D model: similar to the reference model, but with maximum 2R1D interactions allowed for each ray.
    \item 1R1D model: similar to the reference model, but with maximum 1R1D interactions allowed for each ray.
    \item Simplified model: simplified representation of the storage racks as cuboids and maximum 3R1D interactions allowed for each ray. The cuboids are associated with a custom material that allows reflection, diffraction, and transmission.

\end{enumerate}

For the fourth model, default material properties for cuboids were considered first, inspired by the dielectric material properties of wood \cite{ITU} and reasonable but arbitrary transmission losses: 

\begin{itemize}
    \item At 2 GHz: transmission linear loss = 0.4 dB/m; $\varepsilon_\mathrm{r}^{'} = 1.99$ and $\varepsilon_\mathrm{r}^{''} = 0.090$.
    \item At 28 GHz: transmission linear loss = 2.5 dB/m; $\varepsilon_\mathrm{r}^{'} = 1.99$ and $\varepsilon_\mathrm{r}^{''} = 0.11$.
\end{itemize} 

Those values are tuned during the study to better fit the reference case results and obtain a simplified but equivalent material representation. 

Those four propagation models are compared in terms of large-scale channel parameters such as received power, delay spread (DS), horizontal angle-of-arrival spread (HAAS), horizontal angle-of-departure spread (HADS), CTs, and accuracy; for the two considered frequencies.


\section{Simulation results and derived recommendations}

The ray-path data from P2MP simulations for each propagation model is used to calculate and compare the large-scale channel parameters such as received power, DS, HAAS, and HADS for each link; statistical distributions are calculated as well. Then a global comparison is made among different propagation models in terms of CTs and RMSEs.

\subsection{Simulation Results at 2 GHz}

\begin{figure}[ht!] 
\centering
\includegraphics[width=3.3in]{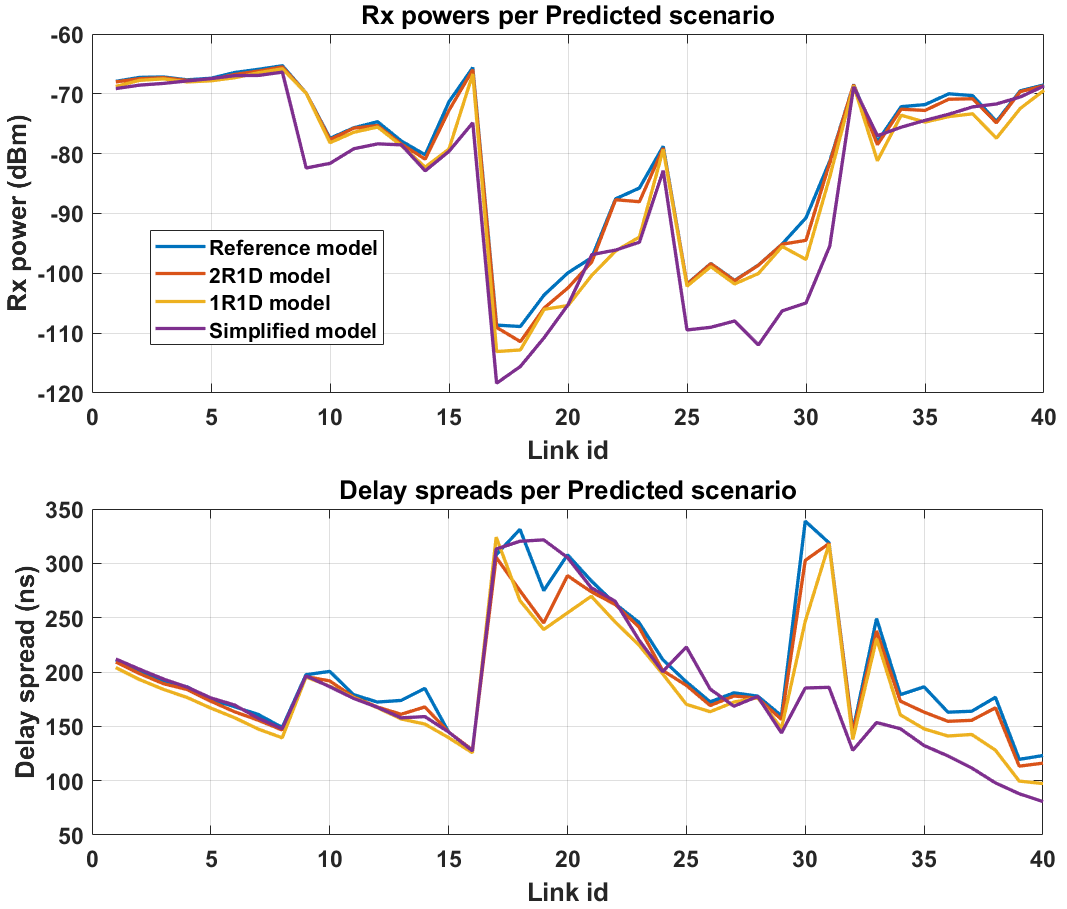}
\caption{Received Power (dBm) and DS (ns) for different links with BS 1 as the transmitter, at 2 GHz.}
\label{5.1}
\end{figure}

Fig. \ref{5.1} shows the received power and DS for all the different links with BS 1 as the transmitter. It can be seen that the second and third propagation models are a close match to the reference case. The fourth propagation model with simplified representation and tuned material properties: transmission linear loss = 0.2 dB/m; $\varepsilon_\mathrm{r}^{'} = 1.2$ and $\varepsilon_\mathrm{r}^{''} = 0.09$, also shows a good match to the reference case for most of the links but varies a lot for some links, for instance: link 30.

\begin{figure}[ht!] 
\centering
\includegraphics[width=3.3in]{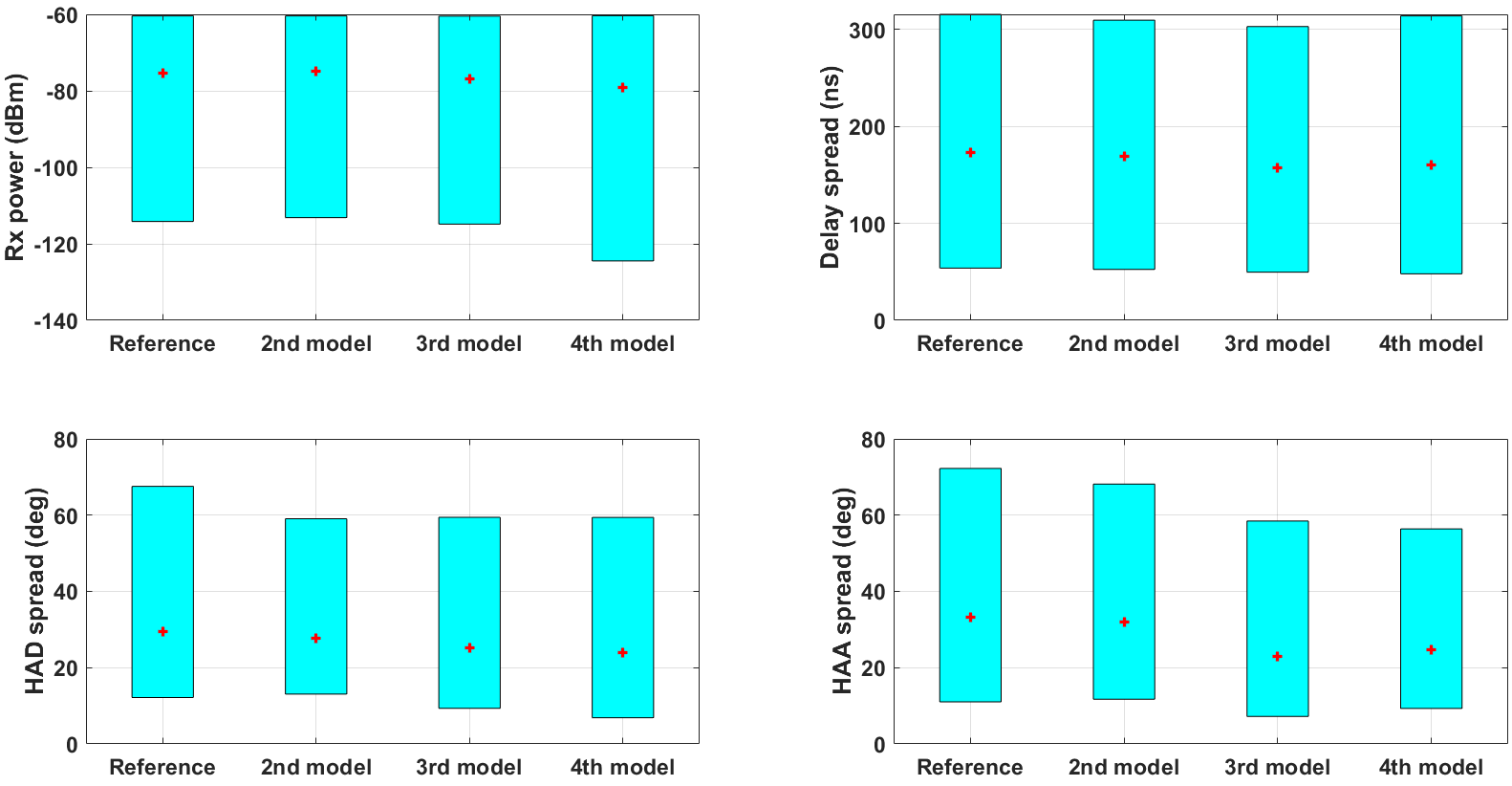}
\caption{Statistical distribution of Received Power, DS, HADS, and HAAS per propagation model, for all possible BS-UT links, at 2 GHz.}
\label{6.1}
\end{figure}

This information alone is not enough to draw any conclusions. Hence, the statistical distribution of received power, DS, HADS, and HAAS for different propagation models for all the possible links between different BS-UT positions shown in Fig. \ref{3.1} is also analyzed. The results are shown in Fig. \ref{6.1}. The red cross corresponds to the median value. The blue area represents the 90\% distribution range, from quantile 5\% to quantile 95\%. The median value of the received power, DS, HADS, and HAAS for the second, third, and fourth propagation models is quite similar to that of the reference model. The distribution range of the received power, DS, HADS, and HAAS for the second model has the closest fit to the reference case. Meanwhile, the distribution range for the third and fourth models varies from the reference case, especially for DS and HAAS. It can be attributed to the simplification of the shapes in the fourth model and quite a low number of allowed interactions in the third model. 

\begin{figure}[ht!] 
\centering
\includegraphics[width=3.2in]{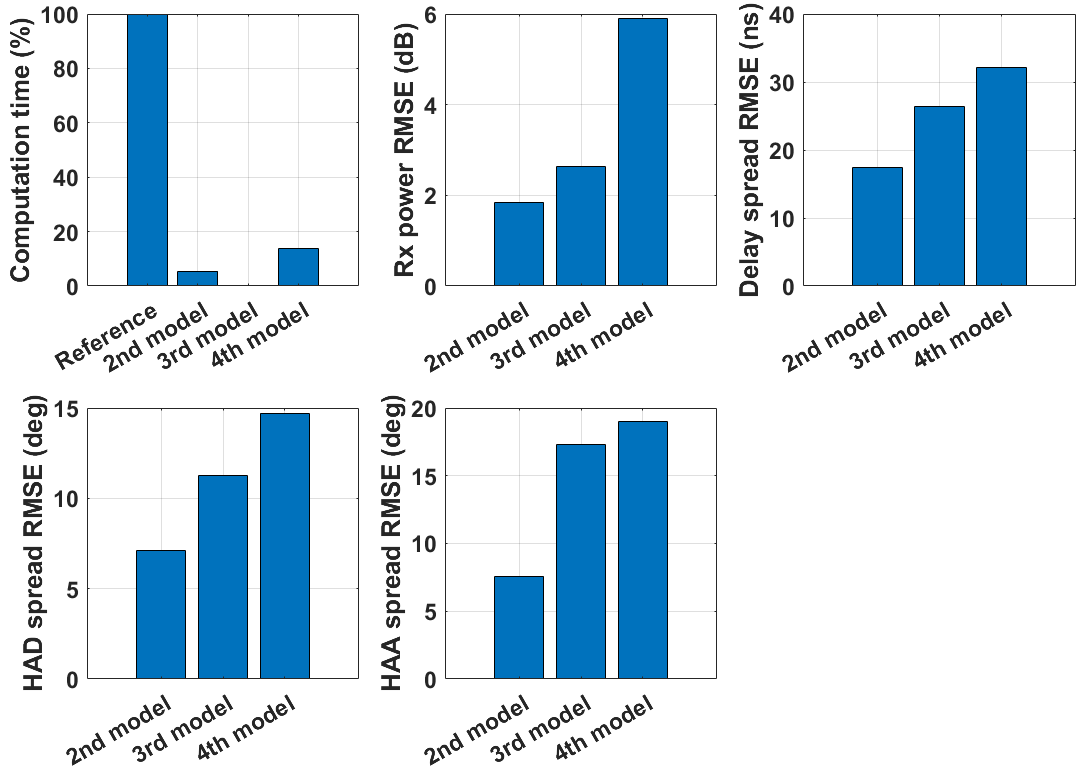}
\caption{CTs and RMSE of Received Power, DS, HADS, and HAAS per propagation model, at 2 GHz.}
\label{7.1}
\end{figure}

In order to conclude, the CTs and RMSE of received power, DS, HADS, and HAAS for different propagation models for all possible BS-UT links are finally analyzed. As shown in Fig. \ref{7.1}, the CT for the reference case is maximum because of the detailed representation and 3R1D allowed for each ray. The CT for the second model is 6\% of the reference CT while it is 16\% of the reference CT for the fourth model. Reduction in computational resource consumption is significant. The RMSE of received power, DS, HADS, and HAAS for each model is calculated with respect to the reference model. Fig. \ref{7.1} clearly shows that the RMSE of received power, DS, HADS, and HAAS is the least for the second model.

\subsection{Simulation Results at 28 GHz}

The analysis is also carried out at a frequency of 28 GHz. The mmWave technology has several potential advantages, including higher bandwidth, lower latency, and the ability to support high data rate transmissions. It is a promising technology for various InF applications, including 5G IIoT. However, at mmWave frequencies, ray-paths are more easily absorbed or blocked by physical objects such as wooden boxes, storage containers, vertical lifts, and other obstructions. It leads to increased attenuation and reduced signal strength. As a result, for this study at 28 GHz, the rays transmitted through the objects play a very small role compared to other interactions (reflections and diffractions) in the prediction of channel parameters. This fact plays a part when the material properties of cuboids in the fourth model are tuned to reduce the RMSEs, at 28 GHz. For the analysis at 28 GHz, BS 4 serves as the transmitter, and twenty-four different UT positions, closest to BS 4, serve as receivers. Most receiver positions are either in LoS or obstructed-LoS (OLoS) condition.

\begin{figure}[ht!] 
\centering
\includegraphics[width=3.3in]{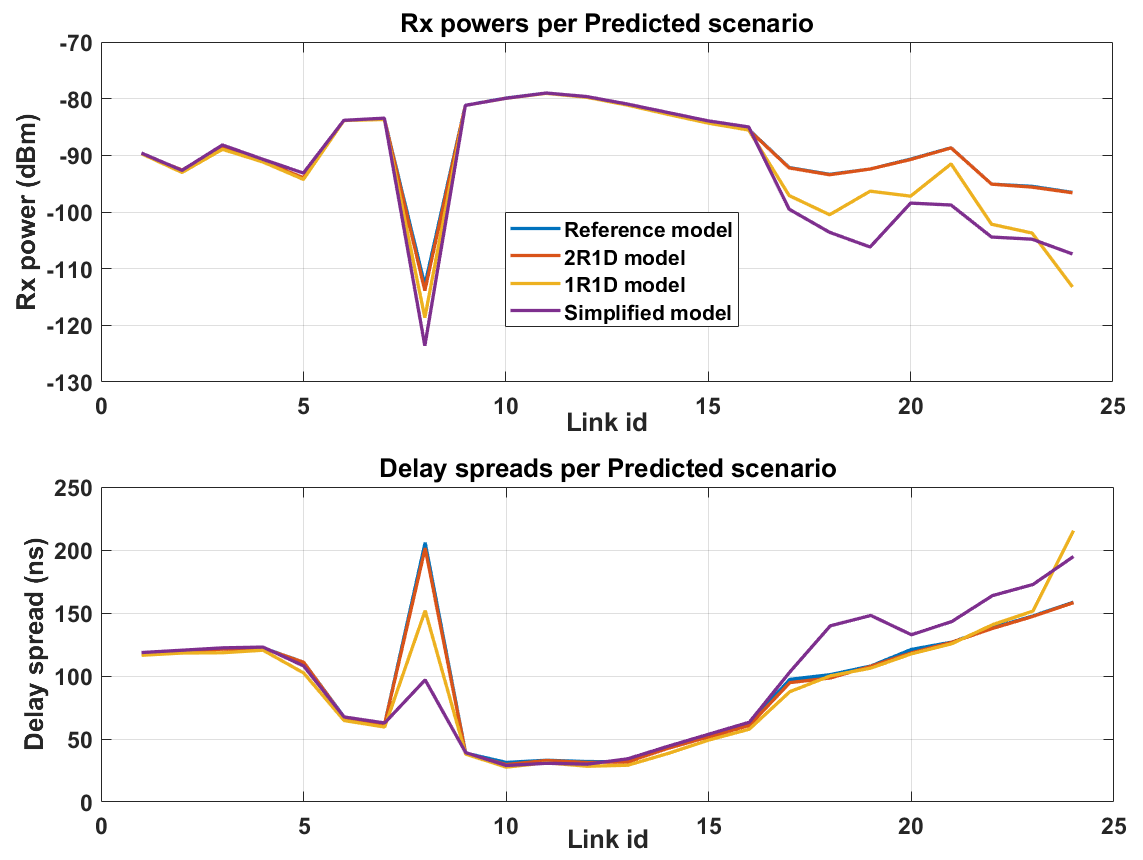}
\caption{Received Power (dBm) and DS (ns) for different links with BS 4 as the transmitter, at 28 GHz.}
\label{10.1}
\end{figure}

Fig. \ref{10.1} shows that, in terms of the received power and DS, the second model is a close match to the reference case. The fourth model with simplified representation and tuned material properties: transmission linear loss = 1.5 dB/m; $\varepsilon_\mathrm{r}^{'} = 2.5$ and $\varepsilon_\mathrm{r}^{''} = 0.11$, also shows a good match to the reference case for most of the links but varies a lot for some links, for instance: link 19, which will subsequently be further discussed.

\begin{figure}[ht!] 
\centering
\includegraphics[width=3.3in]{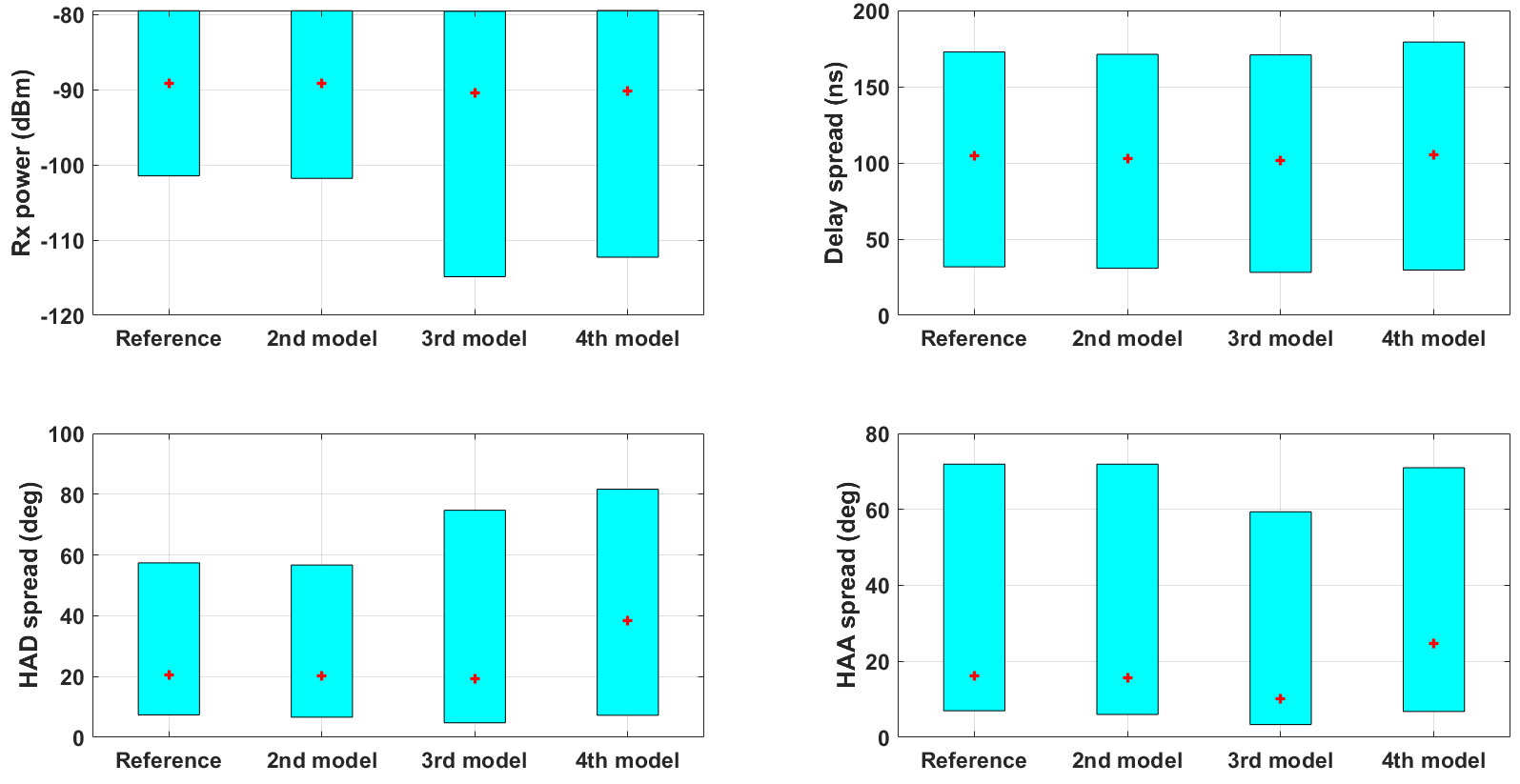}
\caption{Statistical distribution of Received Power, DS, HADS and HAAS per propagation model, for twenty-four links with BS 4 as the transmitter, at 28 GHz.}
\label{8.1}
\end{figure}

The statistical distribution of received power, DS, HADS, and HAAS for twenty-four links with BS 4 as the transmitter for different propagation models is given in Fig. \ref{8.1}. The median value of the received power is around -90 dBm for all the propagation models. The median value of the DS, HADS, and HAAS for the second, third, and fourth propagation models is quite similar to that of the reference model. The distribution range of the received power, DS, HADS, and HAAS for the second model has the closest fit to the reference case. Meanwhile, the distribution range for the third and fourth models varies from the reference case, especially for received power and HADS.

\begin{figure}[ht!] 
\centering
\includegraphics[width=3.3in]{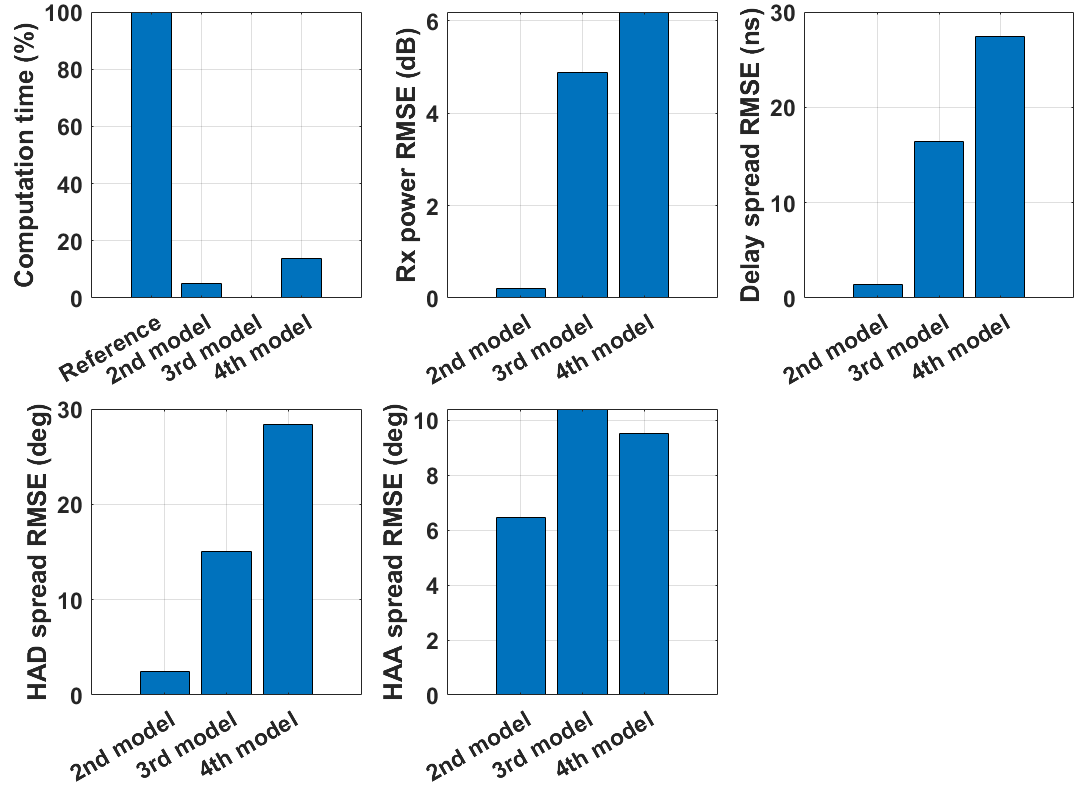}
\caption{CTs and RMSE of Received Power, DS, HADS, and HAAS per propagation model, at 28 GHz.}
\label{9.1}
\end{figure}

As done previously at 2 GHz, in order to conclude at 28 GHz, the CTs and RMSE of received power, DS, HADS, and HAAS are finally analyzed. As shown in Fig. \ref{9.1}, the CT for the reference case is maximum because of the detailed representation and 3R1D allowed for each ray. The CT for the second model is 5\% of the reference CT while it is 14\% of the reference CT for the fourth model. There is a significant CT decrease compared to the reference case. The RMSE of received power, DS, HADS, and HAAS, for each model shown in Fig. \ref{9.1}, is calculated with respect to the reference model. It can be clearly noticed that the RMSE of received power, DS, HADS, and HAAS is the least for the second model with a detailed representation and 2R1D allowed for each ray. 

\subsection{Analysis and Recommendations}

\begin{figure}[ht!] 
\centering
\includegraphics[width=3in]{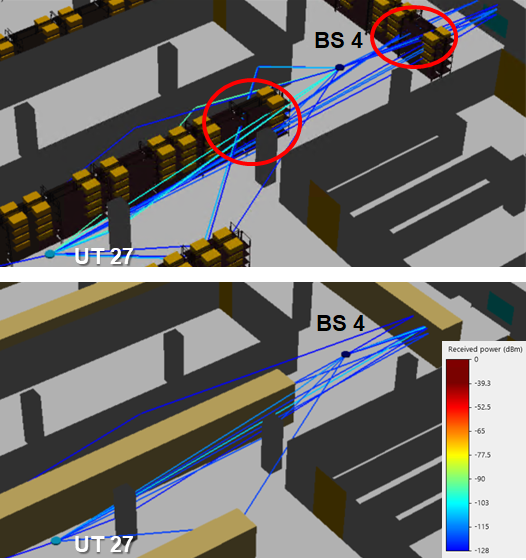}
\caption{Some of the predicted ray paths for link 19, at 28 GHz: Detailed representation (top) vs. Simplified representation (bottom).}
\label{11.3}
\end{figure}

Testing 5G or beyond-5G radio technologies requires realistic propagation channel predictions, encompassing all the complexity and fluctuations that can be encountered in a real environment. The reference model is expected to provide a satisfactory prediction in this sense. In particular, the obtained channel properties on certain links, especially at the higher frequency, do result from strong but very specific propagation paths that pass through the racks without obstruction, for instance, as seen for link 19 (BS 4 - UT 27, at 28 GHz) in Fig. \ref{11.3}. The existence of these paths depends on the exact geometry of the environment; they cannot be reproduced with a simplified representation. 

The fourth model is an acceptable solution for testing purposes, but the second model is the one that offers the best compromise between CTs and accuracy. The RMSE values for the fourth model can be further decreased by fine-tuning the material properties of the cuboids used for simplified representation: at 2 GHz and 28 GHz. The tuned material properties of the cuboids utilized in this study present a substantial improvement over the default material properties in predicting the received power (6 dB reduction in RMSE at 2 GHz), interesting improvement for DS RMSE (5 ns) and HAAS RMSE (2 deg), while the HADS RMSE remains almost unchanged.

Except for a precise factory digital twin that would be continuously updated (which is not feasible yet), the predictions of the three first models are influenced by the environment’s details that may not be an exact replication of reality. Indeed, the packages stored in the racks are moved regularly. In a radio-planning context, where a minimum coverage received power is required, geographical details that change over time cannot be considered. It is preferred to have a pessimistic power estimate such that the required coverage is always guaranteed with a minimum margin. This is achievable with the fourth model, which predicts the received power with reasonable accuracy for most of the links, except when the absence of wooden boxes plays a dominant role in the propagation, especially at higher frequencies. In this situation, the power predicted by the fourth model is pessimistic; but this can be considered a critical defect. Note that the other channel metrics, such as DS, HADS, and HAAS, are usually not considered for radio-planning. 


\section{Conclusion and future perspective}

This work used different propagation models in terms of the number of allowed interactions and simplified representation of storage racks. The propagation model with the highest number of allowed interactions and the most detailed representation was chosen as the reference model. The large-scale channel parameters such as received power, DS, HAAS, HADS, their statistical distribution, CTs, and RMSEs for different propagation models were calculated and compared to the predictions from the reference model. 

The study shows that from the perspective of link- or system-level simulations, a good compromise is reached when employing a lower number of interactions first, such as 2R1D for each ray rather than 3R1D, and using a detailed 3D representation. It will help to significantly reduce the CTs without compromising the accuracy of the predicted channel parameters. One should not significantly decrease the number of allowed interactions as it can lead to inaccurate results. Hence, the second model seems reasonable for producing channel samples devoted to link- or system-level designs. 

On the other hand, from a radio-planning perspective, it is better to use a simplified representation of objects as in the fourth model, still to have significant time savings and no major inaccuracies. Furthermore, the material properties can be fine-tuned to reduce errors. The fourth model provides pessimistic received power for some obstructed links but is related to small details of the 3D model that may not be considered in a radio-planning context. Additionally, if the errors are within an acceptable range, using simplified representations in an InF scenario can ease the process of realizing a 3D digital model of the propagation scenario and lead to significantly lower CTs. 

The facts and recommendations derived from the study presented in this paper can help in enhancing a ray-tracing tool. Moreover, they can be critical for the end-to-end (E2E) workflow of link- and system-level designs, as well as network planning and optimization for 5G smart factories.



\section*{Acknowledgment}

This work is part of a project that has received funding from the European Union's Horizon 2020 research and innovation programme under the Marie Skłodowska Curie grant agreement No. 956670.



\end{document}